\documentclass[12pt,a4paper]{article}
\usepackage[dvips]{graphicx}
\usepackage{amssymb}
\usepackage{rotating}
\makeatletter
\newif\if@preliminary
\@preliminaryfalse
\def\preliminary{\@preliminarytrue}
\def\preprintno#1{\def\@preprintno{#1}}
\def\address#1{\def\@address{#1}}

\def\abstract#1{\def\@abstract{#1}}
\renewcommand\abstractname{\sc Abstract}
\newlength\preprintnoskip
\newlength\abstractwidth
\@titlepagetrue
\renewcommand\maketitle{\begin{titlepage}%
  \let\footnotesize\small
  \def\thefootnote{\fnsymbol{footnote}}
  { \@preprintno  \par}%
   \vspace{0.4cm}
  \begin{center}%
    {\Large\bf\boldmath \@title \par} 
    \vspace{1.5cm}
    {\sc\@author \par}\vskip 6mm%
    {\@address \par}%
    \if@preliminary
      \vskip 2cm {\large\sf PRELIMINARY DRAFT \par \@date}%
    \fi
  \end{center}\par
  \@thanks
   \vspace{0.5cm}
  \begin{center}%
    \parbox{\abstractwidth}{\centerline{\abstractname}%
    \vskip 3mm%
    \@abstract}
  \end{center}
  \end{titlepage}%
  \setcounter{footnote}{0}%
  \def\thefootnote{\arabic{footnote}}
  \let\thanks\relax\let\maketitle\relax
  \gdef\@thanks{}\gdef\@author{}\gdef\@address{}%
  \gdef\@title{}\gdef\@abstract{}\gdef\@preprintno{}
}%
\def\shortletter{%
  \setcounter{secnumdepth}{5}
  \def\paragraph{%
    \@startsection{paragraph}{4}{\parindent}%
      {3.25ex \@plus1ex \@minus.2ex}{-.5em}%
      {\reset@font\normalsize\bfseries}}%
  \renewcommand\theparagraph{\arabic{paragraph}.\hskip-.5em}
  \def\subparagraph{%
    \@startsection{subparagraph}{5}{\parindent}%
      {3.25ex \@plus1ex \@minus.2ex}{-.5em}%
      {\reset@font\normalsize\bfseries}}%
  \renewcommand\thesubparagraph{(\alph{subparagraph})\hskip-.5em}
}
\def\thesection{\arabic{section}.}
\def\appendix{\setcounter{section}{0}
 \def\thesection{Appendix \Alph{section}:}
 \def\theequation{\Alph{section}.\arabic{equation}}}
\evensidemargin=\oddsidemargin  
\def\@citex[#1]#2{\if@filesw\immediate\write\@auxout{\string\citation{#2}}\fi
  \def\@citea{}\@cite{\@for\@citeb:=#2\do
    {\@citea\def\@citea{,\penalty\@m}\@ifundefined
       {b@\@citeb}{{\bf ?}\@warning
       {Citation `\@citeb' on page \thepage \space undefined}}%
\hbox{\csname b@\@citeb\endcsname}}}{#1}}
\def\citerange{\@ifnextchar [{\@tempswatrue\@citexr}{\@tempswafalse\@citexr[]}}
\def\@citexr[#1]#2{\if@filesw\immediate\write\@auxout{\string\citation{#2}}\fi
  \def\@citea{}\@cite{\@for\@citeb:=#2\do
    {\@citea\def\@citea{--\penalty\@m}\@ifundefined
       {b@\@citeb}{{\bf ?}\@warning
       {Citation `\@citeb' on page \thepage \space undefined}}%
\hbox{\csname b@\@citeb\endcsname}}}{#1}}
\long\def\@makecaption#1#2{%
  \vskip\abovecaptionskip
  \sbox\@tempboxa{#1: \emph{#2}}%
  \ifdim \wd\@tempboxa >\hsize
    #1: \emph{#2}\par
  \else
    \hbox to\hsize{\hfil\box\@tempboxa\hfil}%
  \fi
  \vskip\belowcaptionskip}
\def\fmslash{\@ifnextchar[{\fmsl@sh}{\fmsl@sh[0mu]}}
\def\fmsl@sh[#1]#2{%
  \mathchoice
    {\@fmsl@sh\displaystyle{#1}{#2}}%
    {\@fmsl@sh\textstyle{#1}{#2}}%
    {\@fmsl@sh\scriptstyle{#1}{#2}}%
    {\@fmsl@sh\scriptscriptstyle{#1}{#2}}}
\def\@fmsl@sh#1#2#3{\m@th\ooalign{$\hfil#1\mkern#2/\hfil$\crcr$#1#3$}}
\makeatother

\newcounter{actr}

\newcommand{\fnnum}[1]{$^#1$}

\def\today{\number\day
           \space\ifcase\month\or
             January\or February\or March\or April\or May\or June\or
             July\or August\or September\or October\or November\or December\fi
           \space\number\year}
\def\be{\begin{equation}}
\def\ee{\end{equation}}
\def\bea{\begin{eqnarray}}
\def\eea{\end{eqnarray}}

\def\D0{D\O~}

\def\ra{\rightarrow}

\def\to{\rightarrow}

\renewenvironment{thebibliography}[1]
        {\begin{list}{\arabic{enumi}.}
        {\usecounter{enumi}\setlength{\parsep}{0pt}
\setlength{\leftmargin 0.5cm}{\rightmargin 0pt}
         \setlength{\itemsep}{0pt} \settowidth
        {\labelwidth}{#1.}\sloppy}}{\end{list}}
\def\be{\begin{equation}}
\def\ee{\end{equation}}
\def\bea{\begin{eqnarray}}
\def\eea{\end{eqnarray}}

\def\D0{D\O~}

\def\ra{\rightarrow}
\begin{document}
%
\preprintno{\hspace*{-0.65cm}hep-ph/9708402 \hfill  DESY-97-141\\
            August, 1997                   \hfill MSUHEP-70815 }%
\title{%
Probing Electroweak Symmetry Breaking via\\ 
$WW$ Interactions at High Energy Colliders\hspace*{0.1cm}\thanks{%
~Plenary talk presented by C.--P.~Yuan at
``{\it Beyond the Standard Model V} '',  Balholm, Norway, 
    April~29-May~4, 1997; To be published in the conference
    proceedings, 
    Eds. L.~Bugge, P.~Osland and  A.~Read.}
}
\author{
Hong-Jian~He~\fnnum{1},~~~%
Yu-Ping~Kuang~\fnnum2~~~and~~~%
C.--P.~Yuan~\fnnum3 %
}
\address{
\fnnum1{}{\it Theory Division, Deutsches Elektronen-Synchrotron DESY}\\
          {\it D-22603 Hamburg, Germany}\\[0.3cm] 
\fnnum2{}{\it%
           CCAST ( World Laboratory ), P.O.Box 8730, Beijing 100080, China}\\
          {\it Institute of Modern Physics, Tsinghua University,
               Beijing 100084, China\thanks{{~Mailing address.}} }\\[0.3cm]
\fnnum3{}{\it Department of Physics and Astronomy,
            Michigan State University}\\%
          {\it East Lansing, Michigan 48824, USA}
}
\abstract{%
We classify the sensitivities of the future high energy hadron and 
electron colliders, such as the LHC and the future Linear Colliders,
to probing all the next-to-leading order (NLO)
bosonic operators for studying the electroweak symmetry 
breaking (EWSB) mechanism.
We first develop a power counting rule (a generalization of 
Weinberg's counting method for the nonlinear sigma model) 
for the electroweak theories formulated by the chiral Lagrangian. 
Then, we formulate the 
Longitudinal-Goldstone Boson Equivalence Theorem 
as a physical criterion to discriminate the scattering processes
that are not sensitive to the EWSB sector.
The {\it complementarity} of different scattering processes via 
different colliders for a complete probe of all these NLO effective
operators is demonstrated.
\\[0.4cm]
PACS number(s): 11.30.Qc, 11.15.Ex, 12.15.Ji, 14.70.--e \\[0.8cm]
}
\maketitle

\baselineskip20pt  



\section*{\normalsize\bf 1. I\lowercase{ntroduction}}

Probing the electroweak symmetry breaking (EWSB) mechanism is one
of the most outstanding tasks in today's high energy physics.
Before the data is available, it is necessary to study all the possible 
scenarios in which the EWSB sector is either weakly or strongly 
interacting. It is the latter case that we shall discuss in this paper.

At the scale below new heavy resonances, 
the EWSB sector can be parametrized by means of the
electroweak chiral Lagrangian (EWCL) in which the 
$SU(2)_L \otimes U(1)_Y$ gauge symmetry is nonlinearly realized.
Without experimental observation of any
new light resonance in the EWSB sector~\cite{search}, this effective
field theory approach provides the most economic description of
the possible new physics effects 
and is thus {\it complementary} to those specific model 
buildings~\cite{peskin-snow}.
In the present analysis, taking this 
general EWCL approach, we shall concentrate on 
studying the effective bosonic operators among which 
the leading order operators are universal 
and the next-to-leading-order (NLO) operators describe 
the model-dependent new effects.

Following Ref.~\cite{app,etlhc}, the EWCL can be generally written as
\be
{\cal L}_{eff}
= \displaystyle\sum_n 
\ell_n\displaystyle\frac{f_\pi~^{r_n}}{\Lambda^{a_n}}
{\cal O}_n(W_{\mu\nu},B_{\mu\nu},D_\mu U,U,f,\bar{f})
= {\cal L}_G + {\cal L}_{S} + {\cal L}_F \, ,
\label{eq:effL}
\ee
where 
$D_{\mu}U  =  
\partial_{\mu}U + ig{\bf W}_{\mu}U -ig^{\prime}U{\bf B}_{\mu}~$,
${\bf W}_{\mu}\equiv W^a_{\mu}\displaystyle\frac{\tau^a}{2}$,
${\bf B}_{\mu}\equiv B_{\mu}\displaystyle\frac{\tau^3}{2}$, 
$~U  =  \exp [i\tau^a\pi^a/f_\pi ]~$, $\pi^a$ is the Goldstone boson
(GB) field and $f$($\bar{f}$) is the fermion field.
In (\ref{eq:effL}), we have factorized out the dependence on 
$~f_\pi~$ and $~\Lambda~$ so that the dimensionless 
coefficient $\ell_n$  of the operator ${\cal O}_n$ is naturally
of $O(1)$~~\cite{georgi}.
$f_\pi$ ($\simeq 246$\,GeV) is the vacuum expectation value which 
characterizes the EWSB breaking scale.
The effective cut-off scale $\Lambda$ is the 
highest energy scale below which 
(\ref{eq:effL}) is valid. In the case with no new light 
resonance in the EWSB sector,  
$~\Lambda \lesssim 4\pi f_\pi~$~\cite{georgi}.
$~{\cal L}_F~$ is the fermionic part of ${\cal L}_{eff}$.
The bosonic part of the EWCL is given by 
$~{\cal L}_G+{\cal L}_{S}~$ where
${\cal L}_G = -\frac{1}{2}{\rm Tr}({\bf W}_{\mu\nu}{\bf W}^{\mu\nu})
              -\frac{1}{4}B_{\mu\nu}B^{\mu\nu} $ and
$~{\cal L}_S~$ contains operators describing the interactions 
of the gauge bosons ($W^a_\mu$ and $B_\mu$) and the Goldstone bosons. 
Specifically,
\be
{\cal L}_{S}  
 ~=~  {\cal L}^{(2)}+{\cal L}^{(2)\prime}+
             \sum_{n=1}^{14} {\cal L}_n ~,
\label{eq:effL3}
\ee
where ${\cal L}^{(2)}$ is the universal leading order bosonic operator,
and equals to 
$~\left(f_\pi^2/4\right)\times$\linebreak
${\rm Tr}[(D_{\mu}U)^\dagger(D^{\mu}U)]~$.
All the other 15 NLO bosonic operators were 
explicitly given in Refs.~\cite{app,etlhc},
among which
twelve (${\cal L}^{(2)\prime}$ and ${\cal L}_{1\sim 11}$) 
are $CP$-conserving and three 
(${\cal L}_{12\sim 14}$) are $CP$-violating.
If we ignore the small $CP$-violating effects from the 
 Cabibbo-Kobayashi-Maskawa (CKM) mixings in the lowest order 
fermionic Lagrangian $~{\cal L}_{F}~$,
all the one-loop level new divergences generated from 
$~{\cal L}_{\rm G}+{\cal L}_{F}+{\cal L}^{(2)}~$ must be
 $CP$-invariant. 
Therefore, the $CP$-violating operators $~{\cal L}_{12\sim 14}~$,
deduced from some unknown new physics other than the 
CKM mixing, are actually {\it decoupled} at this level, and their
coefficients can have values significantly larger or smaller than that from
the naive dimensional analysis~\cite{georgi}.
Since the true mechanism for $CP$-violation remains un-revealed,  we
shall consider in this paper the coefficients $~\ell_{12\sim 14}~$
to be of $~O(1)~$. We also note that
 the operators $~{\cal L}_{6,7,10}~$ 
violate custodial $SU(2)_C$ symmetry 
(even after $g^{\prime}$ is turned off) in contrast to 
the operators ${\cal L}_{4,5}$ in which the  pure GB interactions
are $SU(2)_C$-invariant. 

The coefficients ($\ell_n$'s) of the 15 NLO operators depend on 
the details of the underlying dynamics. 
Among them, 
$\ell_1$, $\ell_0$ and $\ell_8$ correspond to 
${\rm S}$, ${\rm T}$ and ${\rm U}$
parameters~\cite{app}. ($~{\rm S}=-\ell_1 / \pi$,
$ {\rm T} = \ell_0 /(2\pi e^2)$ and ${\rm U}=-\ell_8 / \pi$.~)
They have been measured 
from the current low energy LEP/SLC data 
and will be further improved at LEPII and upgraded Tevatron.
To distinguish different models of the EWSB mechanism, 
the rest of the $\ell_n$'s  
has to be determined from studying the scattering processes
involving weak gauge bosons. 
The current constraints on the parameters $\ell_{2,3,9}$ and 
$\ell_{4,5,6,7,10}$ from the available collider data at the LEP/SLC 
and the Tevatron are still well above their theoretical natural size
of $O(1)$.\footnote{
Because of limited space, we refer to Ref.~\cite{rev} for details
on these bounds.} This situation makes it extremely desirable to further
test these EWSB parameters at the forthcoming high energy LHC 
and future Linear Colliders (LC)~\cite{lc}.

What is usually done in the literature is to consider only 
a small subset of these operators at a time.
The important question to ask is:
`` How and to what extent can one measure 
{\it all}~ the NLO coefficients $\ell_n$  at the future colliders  
to {\it fully} explore the EWSB sector? ''
To answer this question, one should 
{\bf (i)}. find out, for each given NLO operator, whether 
it can be measured via leading and/or sub-leading amplitudes 
of the relevant processes at each collider;
{\bf (ii)}. determine whether a given NLO 
operator can be sensitively (or marginally sensitively) probed 
through its contributions to the leading (or sub-leading) 
amplitudes of the relevant scattering process at each given collider;
{\bf (iii)}. determine whether carrying out the above study for various 
high energy colliders can {\it complementarily} cover all 
the 15 NLO operators to probe the strongly interacting EWSB sector.
In the following, we show how to meet these {\it Minimal Requirements}
 {\bf (i)}-{\bf (iii)}.

\vspace{0.3cm}
\section*{\normalsize\bf
2. A P\lowercase{ower} C\lowercase{ounting} R\lowercase{ule} 
\lowercase{for} H\lowercase{igh} E\lowercase{nergy} S\lowercase{cattering} 
A\lowercase{mplitudes}}

To make a systematic analysis on the sensitivity of a scattering 
process to probing the new physics operators 
in~(\ref{eq:effL}), we have to first compute the scattering amplitudes 
contributed by those operators. 
For this purpose, we generalize 
Weinberg's power counting rule for the ungauged nonlinear 
sigma model (NLSM)~\cite{wei} and systematically develop a 
power counting method for the EWCL to {\it separately} count the 
power dependences on the energy $E$ and all the relevant mass scales.
Weinberg's counting rule was to count the $E$-power
dependence ($D_E$) for a given $~L$-loop level $S$-matrix element 
$~T~$ in the NLSM. To generalize it to the EWCL, we further include 
the gauge bosons, ghosts, fermions  and possible 
$v^\mu$-factors associated with external 
weak gauge boson ($V= W^\pm , Z^0$) lines [cf.~(\ref{eq:cbv})]. 
After some algebra, we find that for the EWCL and in the energy region 
$~\Lambda > E \gg M_W, m_t~$, 
\be 
D_E = 2L+2+\sum_n {\cal V}_n
\left( d_n+\frac{1}{2}f_n -2\right) -e_v ~,
\label{eq:de}                       
\ee
where ${\cal V}_n$ is the number of type-$n$ vertices in $T$,
$d_n$($f_n$) is the number of derivatives (fermion-lines) 
at a vertex of type-$n$, and
$e_v$ is the number of possible external $v^\mu$-factors
[c.f.~(\ref{eq:cbv})]. 
For external fermions, we consider masses 
$~~m_f\leq m_t \sim O(M_W)\ll E ~$, and the spinor wave functions are 
normalized as 
$~~ \bar{u}(p,s)u(p,s^\prime )=2m_f\delta_{ss^\prime}~$,~ etc.

To correctly estimate the magnitude of each given amplitude $~T~$,
besides counting the power of $E$,
it is also crucial to {\it separately} count the
power dependences on the two typical mass scales of the EWCL: 
the vacuum expectation value $f_\pi$ and the effective cut-off 
$\Lambda$ of the effective theory.
In general, ~$T$~ can always be
written as $~f_\pi^{D_T}~$ times some dimensionless function of 
$~E,~\Lambda$ and $f_\pi$, 
where $D_T = 4-e$ and 
$e$ is the number of external bosonic and fermionic lines.  
Since each of the NLO operators contributing to the
vertices of a Feynman diagram a factor $~1/\Lambda^{a_n}~$,
the total $\Lambda$-dependence in $T$ is
$~1/\Lambda^{\sum_n a_n}~$.~ The power factor $~\Lambda^{a_n}~$
associated with each operator $~{\cal O}_n~$ can be counted by 
the naive dimensional analysis (NDA)~\cite{georgi}.
Bearing in mind the intrinsic $L$-loop factor  
$~(\frac{1}{16\pi^2})^L=(\frac{1}{4\pi})^{2L}~$,  
we can then construct the following precise counting rule for $~T~$
in the energy region $~\Lambda >E \gg M_W, m_t~$:
\bea
T= c_T f_\pi^{D_T}\displaystyle 
\left(\frac{f_\pi}{\Lambda}\right)^{N_{\cal O}}
\left(\frac{E}{f_\pi}\right)^{D_{E0}}
\left(\frac{E}{4 \pi f_\pi}\right)^{D_{EL}}
\left(\frac{M_W}{E}\right)^{e_v} H(\ln E/\mu) ~~,\nonumber \\[0.3cm]
N_{\cal O}=\sum_n a_n~,~~~~ 
\displaystyle D_{E0}=2+\sum_n {\cal V}_n
\left( d_n+\frac{1}{2}f_n-2\right)~, ~~~~ 
D_{EL}=2L~,
\label{eq:counting}
\eea
where the dimensionless coefficient $~c_T~$ contains 
possible powers of gauge couplings ($g,g^\prime$) and Yukawa 
couplings ($y_f$) from the vertices 
of $~T~$, which can be directly counted. $~H$~ is a function 
of $~\ln (E/\mu )~$ coming from loop corrections in the standard
dimensional regularization~\cite{eff}
and is insensitive to $E$.  Neglecting the
insensitive factor $~H(\ln E/\mu)$, we can extract the main features of
scattering amplitudes by simply applying~(\ref{eq:counting}) to
the corresponding Feynman diagrams. 

Note that the counting for $E$-power dependence in 
(\ref{eq:de}) or (\ref{eq:counting})
cannot be directly applied to the amplitudes with external
longitudinal gauge boson ($V_L$) lines. 
Consider the tree-level $V_L V_L \rightarrow V_L V_L$ 
amplitude. Using (\ref{eq:counting}) 
and adding the $E$-factors from the four longitudinal polarization vectors
$~\epsilon_L^{\mu}\sim k^\mu /M_{W,Z}~$, we find that the leading amplitude
is proportional to $E^4/f_\pi^4$ which 
violates the low energy theorem result (i.e. $E^2/f_\pi^2$).
This is because the naive power counting for $V_L$-amplitudes
only gives the leading $E$-power of individual Feynman diagrams, it
does not reflect the fact that gauge invariance causes the 
cancellations of the $E^4$-terms among individual diagrams.
So, how do we count $D_E$ in any amplitude with external $V_L$-lines?
We find that this can be elegantly solved by using the 
Ward-Takahashi (WT) identity \cite{et}:
\be
T[V^{a_1}_L,\cdots ,V^{a_n}_L;\Phi_{\alpha}]
= C\cdot T[-i\pi^{a_1},\cdots ,-i\pi^{a_n};\Phi_{\alpha}]+ B~~,\\
\label{eq:st}
\ee                                               
with\vspace{-0.8cm}
\bea
C \equiv C^{a_1}_{mod}\cdots C^{a_n}_{mod},~
 v^a  \equiv  v^{\mu}V^a_{\mu},~ 
v^{\mu}~\equiv \epsilon^{\mu}_L-k^\mu /M_a 
= O(M_a/E), \nonumber \\ 
B \equiv \sum_{l=1}^n \{~C^{a_{l+1}}_{mod}\cdots C^{a_n}_{mod}
\,T[v^{a_1},\cdots ,v^{a_l},-i\pi^{a_{l+1}},\cdots ,
 -i\pi^{a_n};\Phi_{\alpha}] +{\rm perm.}~\},
\label{eq:cbv}
\eea  
where $\Phi_{\alpha}$ denotes any fields other than
$V^a_L$ or $\pi^a$ in the physical in/out states. 
The constant modification factor
$~C_{mod}^a=1+O({g^2 \over 16 \pi^2})$ in the EWCL
and can be exactly simplified as $1$  
in certain convenient renormalization schemes~\cite{et}. 
Since the right-hand side (RHS)  of (\ref{eq:st}) does not 
have $E$-power cancellations related to external legs,
we can therefore apply our counting rule (\ref{eq:counting}) to 
{\it indirectly count the $D_E$ of
the $V_L$-amplitude via counting the $D_E$ of the RHS of (\ref{eq:st}).}

\vspace{0.3cm}
\section*{\normalsize\bf
3. E\lowercase{stimating} S\lowercase{cattering} A\lowercase{mplitudes and} 
A\lowercase{nalyzing their} 
S\lowercase{ensitivities to} E\lowercase{ach} 
G\lowercase{iven} O\lowercase{perator}}

Using the above counting rule (\ref{eq:counting}), 
we have performed a global analysis for
all $~V^aV^b \to V^cV^d~$ and 
$~f\bar{f}^{(\prime )}\ra V^aV^b,~  
V^aV^bV^c~$ processes
by estimating the contributions from both model-independent operators
(up to one-loop) and the 15 model-dependent NLO operators
(at the tree level)~\cite{etlhc,rev}. 
We reveal a general power counting hierarchy in terms 
of ~$E$, $f_\pi$ and $\Lambda$~ for these amplitudes:

{\small 
\be
\frac{E^2}{f_\pi^2}
\gg \left[\frac{E^2}{f_\pi^2}\frac{E^2}{\Lambda^2}, 
~g\frac{E}{f_\pi}\right] 
\gg \left[g\frac{E}{f_\pi}\frac{E^2}{\Lambda^2}, ~g^2\right] 
\gg \left[g^2\frac{E^2}{\Lambda^2}, ~g^3\frac{f_\pi}{E}\right] 
\gg \left[g^3\frac{Ef_\pi}{\Lambda^2},~g^4\frac{f^2_\pi}{E^2}\right]  
\gg g^4\frac{f_\pi^2}{\Lambda^2} 
\label{eq:pch}
\ee
}
\noindent\noindent
which, in the typical high energy regime  
$~E\in (750\,{\rm GeV},~1.5\,{\rm TeV})$,~ gives
{\small 
$$
\begin{array}{c}
 (9.3,37)
\gg \left[(0.55,8.8),(2.0,4.0)\right]
\gg \left[(0.12,0.93),(0.42,0.42)\right] 
\gg \\
    \left[(0.025,0.099),(0.089,0.045)\right]
\gg \left[(5.3,10.5),(19.0,4.7)\right]\times 10^{-3} 
\gg (1.1,1.1)\times 10^{-3}  
\end{array}
$$
}
\noindent\noindent
where $E$ is taken to be the $VV$-pair invariant mass
and $~\Lambda\approx 4\pi f_\pi\simeq 3.1~$TeV. 
This power counting hierarchy is easy to understand. 
In (\ref{eq:pch}), from left to right, the hierarchy
is built up by increasing either the number of derivatives (i.e. 
power of $E/\Lambda$) or the number of external transverse gauge boson 
$V_T$'s (i.e. the power of gauge couplings).  This power counting
hierarchy provides us a theoretical base to classify all
the relevant scattering amplitudes 
in terms of the three essential parameters $E$, $f_\pi$ and $\Lambda$ 
plus possible gauge/Yukawa coupling constants.
In the high energy region $M_W, m_t\ll E <\Lambda $, and to each order
of chiral perturbation, the leading amplitude for a given scattering 
process is the one with all external $V$-lines being longitudinal, 
and the sub-leading amplitude is the one
with only one external $V_T$-line (and all the other
external $V$-lines being longitudinal). This is because 
the EWCL formalism is a momentum-expansion and 
the GBs are derivatively coupled. 

Using the above power counting rule, we classified in 
Refs.~\cite{etlhc,rev} the most important
leading and sub-leading amplitudes that can probe the
NLO operators via various processes.
That answered the {\it Minimal Requirement}-{\bf (i)}.
To answer the {\it Minimal Requirement}-{\bf (ii)}, we  
shall use the longitudinal-Goldstone boson 
Equivalence Theorem (ET)~\cite{et}
to establish a theoretical criterion for classifying the 
{\it sensitivity} of a given scattering process to each NLO operator.

Let us  consider the scattering process
$W^\pm W^\pm \ra W^\pm W^\pm$ as a typical example to 
illustrate the idea \cite{etlhc,rev}.
The leading and sub-leading amplitudes for this process are given by 
the one with four external $W_L$-lines
(~$T[4W_L]$~) and the one with three external $W_L$-lines plus one 
$W_T$-line (~$T[3W_L,W_T]$~), respectively.
The model-dependent leading contributions in
$~T[4W_L]~$ come from the operators
$~{\cal L}_{4,5}~$. (The contributions from
$~{\cal L}_{2,3,9}~$ in $~T[4W_L]~$ are suppressed by a factor 
$~E^2/f_\pi^2~$ relative to that from $~{\cal L}_{4,5}~$.)
Therefore, it is easier to measure $~{\cal L}_{4,5}~$ than
$~{\cal L}_{2,3,9}~$ via the 
$W^\pm_L W^\pm_L \ra W^\pm_L W^\pm_L$ process.
Furthermore, since the largest contributions in
the sub-leading amplitude $~T[3W_L,W_T]~$ 
come from $~{\cal L}_{3,4,5,9,11,12}~$, this process can be useful for 
probing these operators.
To determine which operators can be sensitively probed via a given 
process, we introduce the following theoretical criterion on the  
sensitivity
of the process to probing a NLO operator.
Consider the contributions of
$~{\cal L}_{4,5}~$ to $~T[4W_L]~$ as an example.
For this case, the WT identity (\ref{eq:st}) gives, 
\be
T[4 W_L]
= C\cdot T[4\pi] + B~~,\\
\label{eq:examp}
\ee                                               
where $~C=1+O({g^2 \over 16 \pi^2})$~, $T[4\pi]=T_0[4\pi]+T_1[4\pi]$
and $B=B^{(0)}_0+B^{(0)}_1$.
($B^{(0)}$ and $B^{(1)}$ 
denote the $B$-term from $V_L$-amplitudes 
with $0$ and $1$ external $V_T$-line, respectively.) 
In the above, ~$T_1[4\pi]$~
contains both the model-independent [~$E^4/(16 \pi^2 f^4_\pi)~$]
and model-dependent contributions 
[~$\ell_{4,5}E^4/(f^2_\pi \Lambda^2)~$];
 $~B^{(0)}_1$  
contains both the model-independent [~$g^2E^2/(16 \pi^2f_\pi^2)~$]
and model-dependent 
[~$\ell_{4,5}~g^2 E^2/\Lambda^2~$]  contributions.
Note that the leading $B$-term $B^{(0)}_0$, which is
of $~O(g^2)$, only depends on the electroweak gauge couplings and
is of the same order as the leading pure 
$W_T$-amplitude $~T[4W_T]~$~\cite{et,etlhc}.
Thus, $B$ {\it is insensitive to the EWSB mechanism}.
To {\it sensitively probe} the EWSB sector by measuring 
$~{\cal L}_{4,5}~$ via $~T[4W_L]~$ amplitude, 
we demand the pure GB-amplitude $~T[4\pi ]~$ contributed from 
$~\ell_{4,5}~$ (as a direct reflection of the EWSB dynamics) 
to dominate over the corresponding model-independent
leading $B$-term (~$B_0^{(0)}~$), i.e.  
~$\ell_{4,5}E^4/(f^2_\pi \Lambda^2) \gg g^2$.  This requirement
builds the {\it equivalence} between the $W_L$'s and GB's amplitudes
in (8), the ET~\cite{etlhc,rev}.                 This gives,
for $~\ell_{4,5}=O(1)~$,   ~$ \frac{1}{4}\frac{E^2}{\Lambda^2}
  \gg \frac{M^2_W}{E^2}~$,
or $~1\gg (0.7 \,{\rm TeV}/E)^4 ~$.
Thus, {\it sensitively probing}  $~{\cal L}_{4,5}~$ 
via the $~4W^\pm_L$-process requires $E\geq 1$\,TeV, 
which agrees with the conclusion from
a detailed Monte Carlo study in Ref.~\cite{wwlhc}.

It is straightforward
to generalize the above discussion to any scattering 
process up to the NLO. In this paper, we classify the sensitivities
of the processes as follows.
For a scattering process involving the NLO coefficient $\ell_n$, if 
$T_1 \gg B~$, then this process is classified to be {\it sensitive} to the 
operator ${\cal L}_n$~.
If not, this process is classified to be 
either {\it marginally sensitive} (for $~T_1 > B~$ but $~T_1 \not\gg B~$)
or {\it insensitive} (for $~T_1 \leq B~$) to the operator ${\cal L}_n$. 
Our results are given in
Table 1, in which {\it both the GB-amplitude and the $B$-term 
are explicitly estimated by our counting rule (\ref{eq:counting}).}
 If $~T_1\leq B~$, this means that the sensitivity is poor so that the
probe of $T_1$ is experimentally harder and requires a higher experimental
precision of at least $O(B)$ to test $T_1$. 
The issue of whether to numerically include $B$
in an explicit calculation of the $V_L$-amplitude is {\it irrelevant}
to the above conclusion.

\vspace{0.3cm}
\section*{\normalsize\bf
4. C\lowercase{lassification} \lowercase{of} S\lowercase{ensitivities} 
\lowercase{to} P\lowercase{robing} EWSB S\lowercase{ector at} 
F\lowercase{uture} H\lowercase{igh} E\lowercase{nergy} 
C\lowercase{olliders} }

This section is devoted to discuss our 
{\it Minimal Requirement}-{\bf (iii)}.
It is understood that the actual sensitivity
of a collider to probing the NLO operators depends
not only on the luminosities of the active partons (including 
weak-gauge bosons)
inside hadrons or electrons (as discussed in Ref.~\cite{etlhc,rev}),
but also on the detection efficiency of the signal events after
applying background-suppressing kinematic cuts
to observe the specific decay mode of the final state
weak-bosons (as discussed in Refs.~\cite{wwlhc,wwnlc}). 
However, all of these will only add fine structures 
to the sub-leading contributions listed in Table~1 but not 
affect our conclusions about the leading contributions
as long as there are enough signal events produced.
This fact was illustrated in Ref.~\cite{etlhc}
for probing the NLO operators via 
$W^\pm W^\pm \ra W^\pm W^\pm$ at the LHC. 
We have further applied the same method to 
other scattering processes (including possible incoming
photon/fermion fields) 
for various high energy colliders with the luminosities of
the active partons included, some of 
the details of the study were given in Ref.~\cite{rev}.\footnote{
For the recent further elaborating numerical analyses on the future 
linear colliders, see Refs.~\cite{hjhe,desy-lc}.} 
In this paper, we shall not perform a detailed numerical study 
like Refs.~\cite{wwlhc,wwnlc},
but only give a first-step qualitative global power counting 
analysis which serves as a useful guideline for our further elaborating
numerical calculations~\cite{hjhe,desy-lc}.

After examining all the relevant $ 2 \ra 2$ and $2 \ra 3$ hard scattering
processes at the LHC and the LC,
we summarize~\cite{rev,japan} in Table~1 our global classification for 
the sensitivities of various 
future high energy colliders to probing the 15 
model-dependent NLO bosonic operators.
Here, the energy-$E$ represents the typical energy scale
of the hard scattering processes under consideration.
The leading $B$-term for each high energy
process is also listed and compared with the corresponding
$V_L$-amplitude. If the polarizations of the 
initial/final state gauge bosons are not distinguished 
but simply summed up, the largest $B$ 
in each process (including all possible polarization states)
should be considered for comparison. 
[If the leading $B_0$, with just one $v_\mu$-factor, cf. Eq.~(6), 
 happens to be zero, then the largest next-to-leading
 term, either the part of $B_0$ term that contains
two (or more) $v_\mu$-factors or the $B_1$ term, should be considered. 
Examples are the $~ZZ\rightarrow ZZ~$ and
$~f\bar{f}\rightarrow ZZZ~$ processes.]
By comparing $T_1$ with $B$ in Table~1 and applying our theoretical 
criterion for classifying the sensitivities, we find that
for the typical energy scale ($E$) of the relevant processes at each
collider, the leading contributions (~marked by $\surd~$) 
can be sensitively probed, while the sub-leading contributions
(~marked by $\triangle~$) can only be marginally sensitively 
probed.\footnote{The exceptions are 
$~f\bar{f}^{(\prime )}\rightarrow W^+W^-/(LT),W^\pm Z/(LT)~$ 
for which $~T_1 \leq B_0$~. Thus the probe of them is insensitive.
($L/T$ denotes the longitudinal/transverse polarizations of
$~W^\pm ,~Z^0~$ bosons.) }
(To save space, 
Table~1 does not list those processes to which the NLO operators 
contribute {\it only} sub-leading amplitudes. These processes
are $~WW\rightarrow W\gamma ,Z\gamma +{\rm perm.}~$ 
and $~f\bar{f}^{(\prime )}\rightarrow W\gamma ,WW\gamma , WZ\gamma 
~$, which all have one external transverse $\gamma$-line and 
are at most marginally sensitive.)

From Table~1, some of our conclusions can be drawn as follows. \\
~~~{\bf (1).}
At LC(0.5), which is a LC with $\sqrt{S}=0.5$\,TeV, $\ell_{2,3,9}$
can be sensitively probed via $e^-e^+ \rightarrow W^-_L W^+_L$. \\ 
~~~{\bf (2).}
For pure $V_L V_L \rightarrow V_L V_L$ scattering amplitudes, 
the model-dependent operators ${\cal L}_{4,5}$
and ${\cal L}_{6,7}$ can be probed 
most sensitively. 
 ${\ell}_{10}$ can only be sensitively probed 
via the scattering process $Z_LZ_L \rightarrow Z_LZ_L$ which 
is easier to detect at the LC(1.5) [a $e^-e^+$ or $e^-e^-$ collider 
with $\sqrt{S}=1.5$\,TeV] than at the LHC(14) [a pp collider with
$\sqrt{S}=14$\,TeV].\footnote{
Studying $~f\bar{f}^{(\prime )}\rightarrow V^aV^bV^c$ 
can provide complementary information 
on the operators ${\cal L}_{4,5,6,7,10}$~\cite{hjhe}.
} 
\\ 
~~~{\bf (3).}
The contributions from ${\cal L}^{(2)\prime}$~ and 
${\cal L}_{2,3,9}$ to the pure $4V_L$-scattering processes
 lose the $E$-power dependence by a 
factor of $2$. Hence, the pure $4V_L$-channel is 
less sensitive to these operators. 
[Note that ${\cal L}_{2,3,9}$  can be sensitively 
probed via $f {\bar f} \rightarrow W_L^-W_L^+$ process at LC(0.5) and LHC(14).]
The pure $4V_L$-channel cannot probe ${\cal L}_{1,8,11\sim 14}$ which
can only be probed via processes with $V_T$('s). 
Among ${\cal L}_{1,8,11\sim 14}$,
the contributions from $~{\cal L}_{11,12}~$ to processes 
with $V_T$('s) are most important, although their contributions 
are relatively suppressed by a factor $gf_\pi /E$  as compared to
the leading contributions from 
${\cal L}_{4,5}$ to pure $4V_L$-scatterings.
${\cal L}_{1,8,13,14}$ are generally suppressed by higher powers of
$gf_\pi /E$ and are thus the least sensitive.
The above conclusions hold for both LHC(14) 
and LC(1.5). \\
~~~{\bf (4).} 
At LHC(14), ${\ell}_{11,12}$ 
can be sensitively probed via $q \bar q' \rightarrow W^\pm Z$
whose final state is not electrically neutral. Thus, 
this final state is not accessible at LC. 
Hence, LC(0.5) will not be sensitive to these operators.
To sensitively probe ${\ell}_{11,12}$ at LC(1.5), one has to measure
$e^-e^+ \rightarrow W^-_L W^+_L Z_L$. 
\\ 
~~~{\bf (5).}
To sensitively probe ${\ell}_{13,14}$,
a high energy $~e^-\gamma~$ linear collider 
is needed for studying the processes 
$~e^-\gamma \rightarrow \nu_e W^-_LZ_L,~e^-W^-_LW^+_L~$, 
in which the backgrounds \cite{eAback}~ are much
less severe than those for $\gamma \gamma \rightarrow W^+_L W^-_L$,
 whose  amplitude is of the order of 
$~e^2\frac{E^2}{\Lambda^2}~$, to which the operators ${\cal L}_{13,14}$
(and also ${\cal L}_{1,2,3,8,9}$) can contribute. Thus, the latter process
would be useful for probing ${\ell}_{13,14}$ 
at a $\gamma\gamma$ collider
if the backgrounds could be efficiently suppressed.

We also note that to measure the individual coefficient of the 
NLO operator, one has to be able to separate, for example, the
$W^+W^- \ra Z^0 Z^0$ and the $Z^0Z^0 \ra Z^0 Z^0$ scattering processes.
Although this task can be easily done at the LC by tagging the forward
leptons,\footnote{Note also that the accidental cancellation in the 
$e$-$e$-$Z$ vector coupling makes the rate of  
$e^-e^+\to e^-e^+Z^0Z^0~ (Z^0Z^0 \ra Z^0 Z^0)$
rather small in comparison with that of 
$e^-e^+\to \nu\bar{\nu}W^-W^+~ (W^+W^- \ra Z^0 Z^0)$ at the LC.
}~ 
it will be a great challenge at the LHC because 
both the up- and down-type quarks from the initial state contribute to the 
scattering processes. Another difficulty for doing the above measurement 
at the LHC is that the hadronic mode of the final state is
unlikely to be useful and the clean leptonic mode has a very small branching 
ratio. Hence, further elaborating numerical analyses would be desirable.

From the above conclusions,  we speculate that 
if there is no new resonance much below the TeV scale and 
the coefficients of the NLO 
operators are not well above the natural size suggested by 
the naive dimensional analysis~\cite{georgi}, the LHC alone may not be 
able to sensitively measure all these operators before 
accumulating a much higher integrated luminosity,
and the linear colliders (LC) are needed to {\it complementarily}
cover the rest of the NLO operators.
In fact, the different phases of 500~GeV and 1.5~TeV energies at the LC
are necessary because they will be sensitive to different 
NLO operators.  An electron-photon (or a photon-photon) 
collider is also very useful in measuring some NLO operators
for achieving a {\it complete} understanding of the underlying
strong EWSB dynamics. \\[1cm]

\vspace{0.3cm}
\section*{\normalsize\bf  A\lowercase{cknowledgements}}
\vspace{-0.1cm}
C.P.Y. thanks Per Osland for invitation and warm hospitality.
We are grateful to many colleagues, especially
Michael Chanowitz, John Donoghue, Tao Han and Peter Zerwas, for useful
discussions on this subject.
H.J.H. is supported by the AvH of Germany;
Y.P.K. by the NSF of China
and the FRF of Tsinghua University; and
C.P.Y. acknowledges the NSF grant under contract PHY-9507683.

\vspace{0.3cm}
\section*{\normalsize\bf  R\lowercase{eferences}}
\vspace{-0.1cm}



\addtolength{\textheight}{1cm}
\newpage

\tabcolsep 1pt
\begin{table}[H] 
{\bf Table 1.}~~~~~ 
  Probing the EWSB Sector at High Energy Colliders:\\[-0.7cm]
\begin{center}
   A Global Classification for the next-to-leading order
   Bosonic Operators \\[0.3cm]
\end{center} 
 {\small (~Notations: ~$\surd =~$Leading contributions, 
 $~\triangle =~$Sub-leading contributions,~
 and ~$\bot =~$Low-energy contributions.~
 ~Notes:~ $^{\dagger}$Here, $~{\cal L}_{13}$ or $~{\cal L}_{14}~$
 does not contribute at $~O(1/\Lambda^2)~$.   ~~ $^\ddagger$At LHC($14$),  
 $W^+W^+\rightarrow W^+W^+$ should also be included.~)}\\[0.3cm] 
\vspace{0.3cm}
\begin{sideways}
\small
\begin{tabular}{||c||c|c|c|c|c|c|c|c|c|c||c||c||} 
\hline\hline
& & & & & & & & & & & & \\
 Operators 
& $ {\cal L}^{(2)\prime} $ 
& $ {\cal L}_{1,13} $ 
& $ {\cal L}_2 $
& $ {\cal L}_3 $
& $ {\cal L}_{4,5} $
& $ {\cal L}_{6,7} $ 
& $ {\cal L}_{8,14} $ 
& $ {\cal L}_{9} $
& $ {\cal L}_{10} $
& $ {\cal L}_{11,12} $
& $T_1~\parallel  ~B$ 
& Processes \\
& & & & & & & & & & & & \\
\hline\hline
 LEP-I (S,T,U) 
& $\bot$ 
& $\bot~^\dagger$
&  
& 
& 
& 
& $\bot~^\dagger$
& 
&
&
& $g^4\frac{f^2_\pi}{\Lambda^2}$ 
& $e^-e^+\ra Z \ra f\bar{f}$\\ 
\hline
  LEP-II
& $\bot$ 
& $\bot$  
& $\bot$  
& $\bot$  
&  
& 
& $\bot$  
& $\bot$ 
&
& $\bot$  
& $g^4\frac{f^2_\pi}{\Lambda^2}$
& $e^-e^+ \ra W^-W^+$\\
\hline
  LC($0.5$)/LHC($14$)
& 
& 
& $\surd$
& $\surd$
& 
& 
& 
& $\surd$
&
& 
& $g^2\frac{E^2}{\Lambda^2} \parallel g^2\frac{M_W^2}{E^2}$
& $f \bar f\ra W^-W^+ /(LL)$\\  
& 
& $\triangle$
& $\triangle$
& $\triangle$
& 
& 
& $\triangle$
& $\triangle$
&
& $\triangle$
& $g^3\frac{Ef_\pi}{\Lambda^2} \parallel g^2\frac{M_W}{E} $ 
& $f \bar f\ra W^-W^+/(LT) $\\  
\hline
& 
& 
& 
& $\surd$
& $\surd$
& $\surd$
& 
& $\surd$
&
& $\surd$
& $g^2\frac{1}{f_\pi}\frac{E^2}{\Lambda^2}
  \| g^3\frac{M_W}{E^2} $
& $f \bar f\ra W^-W^+Z /(LLL) $\\
& 
& $\triangle$ 
& $\triangle$
& $\triangle$
& $\triangle$
& $\triangle$
& $\triangle$
& $\triangle$
&
& $\triangle$
& $g^3\frac{E}{\Lambda^2}\parallel g^3\frac{M_W^2}{E^3}$  
& $f \bar f\ra W^- W^+ Z /(LLT)  $\\
& 
& 
&  
& $\surd$
& $\surd$
& $\surd$
& 
& 
& $\surd$
& 
& $g^2\frac{1}{f_\pi}\frac{E^2}{\Lambda^2}\parallel 
  g^3\frac{M_W}{\Lambda^2}$
& $f \bar f \ra ZZZ /(LLL) $\\
& 
& 
& 
& 
& $\triangle$
& $\triangle$
& 
& 
& $\triangle$
& 
& $g^3\frac{E}{\Lambda^2}\parallel
   g^3\frac{f_\pi}{\Lambda^2}\frac{M_W}{E}$ 
& $f \bar f \ra ZZZ  /(LLT)  $\\
 ~LC($1.0~\&~1.5$)~ 
& 
& 
& 
& 
& $\surd$
& 
& 
&
& 
& 
& $\frac{E^2}{f_\pi^2}\frac{E^2}{\Lambda^2}\parallel g^2$ 
& $W^-W^\pm \ra W^-W^\pm /(LLLL)~^\ddagger$\\
 $\&$ LHC($14$) 
&
& 
& 
& $\triangle$
& $\triangle$
& 
& 
& $\triangle$
&
& $\triangle$
& $g\frac{E}{f_\pi}\frac{E^2}{\Lambda^2}\parallel g^2\frac{M_W}{E}$ 
& $W^-W^\pm\ra W^-W^\pm /(LLLT)~^\ddagger$ \\
& 
& 
& 
& 
& $\surd$
& $\surd$
&
& 
&
& 
& $\frac{E^2}{f_\pi^2}\frac{E^2}{\Lambda^2}\parallel g^2 $
& $W^-W^+ \ra ZZ ~\&~{\rm perm.}/(LLLL)$ \\
&
& 
& $\triangle$
& $\triangle$
& $\triangle$
& $\triangle$
& 
& $\triangle$
&
& $\triangle$
& $g\frac{E}{f_\pi}\frac{E^2}{\Lambda^2}\parallel g^2\frac{M_W}{E}$ 
& $W^-W^+ \ra ZZ ~\&~{\rm perm.} /(LLLT)$ \\
& 
& 
& 
& 
& $\surd$
& $\surd$
& 
& 
& $\surd$ 
&  
& $\frac{E^2}{f_\pi^2}\frac{E^2}{\Lambda^2}\parallel
   g^2\frac{E^2}{\Lambda^2} $
& $ZZ\ra ZZ /(LLLL) $\\
&
& 
& 
& $\triangle$
& $\triangle$
& $\triangle$
&  
&
& $\triangle$
&
& $g\frac{E}{f_\pi}\frac{E^2}{\Lambda^2}\parallel
  g^2\frac{M_WE}{\Lambda^2}$ 
& $ZZ\ra ZZ /(LLLT) $\\
\hline
& 
& 
& 
& $\surd$
& 
& 
& 
&
& 
& $\surd$
& $g^2\frac{E^2}{\Lambda^2} \parallel  g^2\frac{M^2_W}{E^2}$
& $q\bar{q'}\ra W^\pm Z /(LL) $\\
& 
& $\triangle$
& $\triangle$
& $\triangle$
& 
& 
& $\triangle$
& $\triangle$
&
& $\triangle$
& $g^3\frac{Ef_\pi}{\Lambda^2}\parallel g^2\frac{M_W}{E}$ 
& $q\bar{q'}\ra W^\pm Z /(LT) $\\
 LHC($14$)
& 
& 
& 
& $\surd$
& $\surd$
& 
& 
& $\surd$
&
& $\surd$
& $g^2\frac{1}{f_\pi}\frac{E^2}{\Lambda^2}\parallel g^3\frac{M_W}{E^2}$
& $q \bar{q'}\ra W^-W^+W^\pm /(LLL) $\\
& 
& 
& $\triangle$
& $\triangle$
& $\triangle$
&
& $\triangle$
& $\triangle$
&
& $\triangle$
& $g^3\frac{E}{\Lambda^2}\parallel g^3\frac{M_W^2}{E^3}$ 
& $q \bar{q'} \ra W^- W^+W^\pm  /(LLT)  $\\
& 
& 
& 
& $\surd$
& $\surd$
& $\surd$
& 
& 
&
& $\surd$
& $g^2\frac{1}{f_\pi}\frac{E^2}{\Lambda^2}\parallel g^3\frac{M_W}{E^2}$
& $q \bar{q'}\ra W^\pm ZZ /(LLL) $\\
& 
& $\triangle$
& $\triangle$
& $\triangle$
& $\triangle$
& $\triangle$
& $\triangle$
& $\triangle$
&
& $\triangle$
& $g^3\frac{E}{\Lambda^2}\parallel  g^3\frac{M_W^2}{E^3}$ 
& $q \bar{q'} \ra W^\pm ZZ  /(LLT)  $\\
\hline
LC$(e^-\gamma )$
& 
& $\surd$
& $\surd$
& $\surd$
& 
& 
& $\surd$
& $\surd$
&
& $\surd$
& $eg^2\frac{E}{\Lambda^2}\parallel eg^2\frac{M_W^2}{E^3}$ 
& $e^-\gamma \ra \nu_e W^-Z,e^-WW /(LL)$\\
\hline
& 
& $\surd$
& $\surd$
& $\surd$
& 
& 
& $\surd$
& $\surd$
&
& 
& $e^2\frac{E^2}{\Lambda^2}\parallel e^2\frac{M_W^2}{E^2}$ 
& $\gamma \gamma \ra W^- W^+ /(LL)$\\
LC($\gamma \gamma $)
& 
& $\triangle$
& $\triangle$
& $\triangle$
& 
& 
& $\triangle$
& $\triangle$
&
& 
& $e^2g\frac{Ef_\pi}{\Lambda^2}\parallel e^2\frac{M_W}{E}$  
& $\gamma\gamma \ra W^-W^+ /(LT)$\\
& & & & & & & & & & & & \\
\hline\hline 
\end{tabular}
\end{sideways}
\end{table}

 

\begin{thebibliography}{99}

\bibitem{search}
B.C.~Allanach et al, {\it Report of the Working Group on `Searches',}
hep-ph/9708250 (August, 1997), and references therein.

\bibitem{peskin-snow}
E.g., M.~Peskin, Talk at the Snowmass Conference (June, 1996), and the
working group summary report, hep-ph/9704217.

\bibitem{app}
T.~Appelquist et al, Phys. Rev. {\bf D48} (1993) 3235;
and references therein.

\bibitem{etlhc}
H.-J.~He, Y.-P.~Kuang, and C.-P.~Yuan, Phys.~Rev. {\bf D55} (1997) 3038,
Mod. Phys. Lett. {\bf A11} (1996) 3061. 

\bibitem{georgi}
H.~Georgi, {\it Weak Interaction and Modern Particle Theory,} 
Benjamin Pub., 1984.

\bibitem{rev}
H.-J.~He, Y.-P.~Kuang, and C.-P.~Yuan, hep-ph/9704276 and DESY-97-056,
Lectures in the Proceedings of the CCAST (World Laboratory) Workshop
on {\it Physics at TeV Energy Scale,} Vol.~72, pp.119-234.

\bibitem{lc}
E.~Accomando et al, {\it Physics with $e^+e^-$ Linear Colliders,}
(ECFA/DESY LC Physics Working Group),
DESY-97-100 and hep-ph/9705442; H.~Murayama and M.~Peskin,
Ann. Rev. Nucl. $\&$ Part. Sci. {\bf 46} (1996) 533.

\bibitem{wei}
S.~Weinberg, Physica {\bf 96A} (1979) 327.

\bibitem{eff}
For a nice review, 
H.~Georgi, Ann. Rev. Nucl. $\&$ Part. Sci. {\bf 43} (1994) 209.

\bibitem{et}
H.-J.~He, Y.-P.~Kuang, and C.-P.~Yuan, 
Phys. Rev. {\bf D51} (1995) 6463; 
H.-J.~He and W.B.~Kilgore, Phys. Rev. {\bf D55} (1997) 1515; 
H.-J.~He, Y.-P.~Kuang, and X.~Li, Phys. Rev. Lett. {\bf 69} (1992) 2619;
Phys. Rev. {\bf D49} (1994) 4842; Phys. Lett. {\bf B329} (1994) 278;
and references therein.

\bibitem{wwlhc}
J.~Bagger, V.~Barger, K.~Cheung, J.~Gunion, T.~Han,
G.A.~Ladinsky, R.~Rosenfeld, and C.-P.~Yuan,
Phys. Rev. {\bf D49} (1994) 1246; ~{\bf D52} (1995) 3878. 

\bibitem{wwnlc}
 V.~Barger, K.~Cheung, T.~Han, and R.J.N.~Phillips, 
Phys. Rev. {\bf D52} (1995) 3815.

\bibitem{hjhe}
H.-J.~He, DESY-97-140, in these proceedings.

\bibitem{desy-lc}
E.~Boos, H.-J.~He, W.~Kilian, A.~Pukhov, C.-P.~Yuan, and P.M.~Zerwas,
hep-ph/9708310 and DESY-96-256, August, 1997.

\bibitem{japan}
H.-J.~He, Y.-P.~Kuang, and C.-P.~Yuan, Phys. Lett. {\bf B382} (1996) 149. 

\bibitem{eAback}
K.~Cheung, S.~Dawson, T.~Han, and
G.~Valencia, Phys. Rev. {\bf D51} (1995) 5.

\end{thebibliography}
\end{document}
\end